**Interlayer Fermi Polarons in Bilayer $MoTe_2$**

Ruihao Ni[1], Eugen Dizer[2], Maximilian Wolf[2], Son T. Le[3,4], Sharadh Jois[3,4], Jeffrey J. Schwartz[3,4], Liuxin Gu[1], Rundong Ma[1], Suji Park[5], Beini Gao[6], Lifu Zhang[1], Houk Jang[5], Takashi Taniguchi[7], Kenji Watanabe[8], Aubrey T. Hanbicki[3], Adam L. Friedman[3], Richard Schmidt[2], You Zhou[1,9]

[1]Department of Materials Science and Engineering, University of Maryland, College Park, MD 20742, USA

[2]Institut für Theoretische Physik, Universität Heidelberg, 69120 Heidelberg, Germany

[3]Laboratory for Physical Sciences, 8050 Greenmead Drive, College Park, MD 20740, USA

[4]Department of Electrical and Computer Engineering, University of Maryland, College Park, MD, 20742, USA

[5]Center for Functional Nanomaterials, Brookhaven National Laboratory, Upton, NY 11973, USA

[6]Department of Physics, University of Maryland, College Park, MD 20742, USA

[7]Research Center for Electronic and Optical Materials, National Institute for Materials Science, 11 Namiki, Tsukuba 305-0044, Japan

[8]Research Center for Materials Nanoarchitectonics, National Institute for Materials Science, 1-1 Namiki, Tsukuba 305-0044, Japan

[9]Maryland Quantum Materials Center, Department of Physics, University of Maryland, College Park, MD 20742 USA

To whom correspondence should be addressed: youzhou@umd.edu

Atomic bilayers of transition metal dichalcogenides (TMDs) host quantum phases governed by the layer degree of freedom, including bilayer Wigner crystals, fractional Chern insulators, and exciton condensates. These phases are probed primarily through exciton spectroscopy, yet it remains poorly understood how excitons and carriers interact to form Fermi polarons in bilayers, where both the impurity and the Fermi sea carry a layer pseudospin. Progress has been limited because most TMD bilayers have momentum-indirect optical bandgaps, in which non-radiative decay and inhomogeneous broadening obscure the intrinsic spectra. Here, we show that bilayer $MoTe_2$, unlike most TMD bilayers, retains a direct optical bandgap, providing a clean platform for studying bilayer Fermi-polaron physics. In a dual-gated device, an out-of-plane electric field continuously tunes the hybridization between intralayer and interlayer excitons, forming layer-coherent excitons. Upon electrostatic doping, the excitonic spectrum evolves into multiple polaron branches, controlled by both carrier doping and the out-of-plane electric field. Among these, we identify a polaron with no analog in monolayers, in which a layer-coherent exciton is dressed by carriers in the opposite layer, and is quantitatively captured by our field-theoretic model. Our results establish that pseudospin structure in both the impurity and the bath reshapes polaron formation**,** opening new avenues to many-body states such as Bose-Einstein condensates with interlayer coherence.

## I. INTRODUCTION

An impurity immersed in a quantum bath is a canonical many-body problem, realized in systems ranging from ultracold atomic gases to doped semiconductors[1-7]. In atomically thin transition-metal dichalcogenides (TMDs), an optically injected exciton acts as the impurity and interacts strongly with the Fermi sea of free carriers to form attractive and repulsive Fermi polarons[8-12], a picture now well established in monolayers.

Adding a second layer fundamentally changes the problem. In TMD bilayers, both the exciton constituents and the Fermi sea acquire a layer pseudospin, with the upper and lower layers acting as two pseudospin states[13-15]. Because impurity and bath each carry this internal degree of freedom, additional interaction pathways open up through interlayer tunneling, Coulomb interactions, and exciton hybridization[16-18], making the polaron problem both distinct from conventional Fermi-polaron systems and considerably harder to treat theoretically.

Experimentally, elucidating polarons in bilayers is critical because exciton spectroscopy has served as the primary probe of quantum phenomena related to the layer pseudospin, including interlayer coherence[19-21], integer and fractional Chern insulators[22-27], and exciton condensation and superfluidity[28-32]. However, probing polaron physics experimentally has also been difficult: in most TMD homo- and hetero-bilayers, the lowest-energy excitons are momentum-indirect, leading to weak absorption and phonon-assisted emission[19,20,33-35], so intrinsic polaron line shapes are easily masked by extrinsic broadening such as non-radiative decay[36-38].

Here, we show that $MoTe_2$ homobilayers overcome these challenges. Using gate-dependent reflectance and photoluminescence spectroscopy, we identify a direct optical bandgap with strong absorption from K-valley interlayer excitons. In the charge-neutral regime, an out-of-plane electric field controls interlayer–intralayer exciton hybridization originating from tunneling. Intriguingly, upon electrostatic doping, the excitonic spectrum evolves into multiple field- and carrier-density-tunable polaron branches, including a distinct interlayer polaron. This branch originates from a layer-polarized Fermi sea dressing a hybridized intralayer exciton in the opposite layer, and is quantitatively reproduced by a field-theoretic model whose couplings are fixed by independently measured tunnelling and trion binding energies. These results identify a class of Fermi polaron with no monolayer analog, and open exciting avenues for probing and controlling many-body quantum states related to layer pseudospin.

## II. INTRALAYER AND INTERLAYER EXCITON HYBRIDIZATION IN A DIRECT-BANDGAP BILAYER

In our experiment, we encapsulate an exfoliated natural 2H-$MoTe_2$ bilayer between hexagonal boron nitride (hBN) to form a dual-gated device with independent control of the displacement field and carrier density[39-41] (**Fig. 1a**). At zero gate bias, the $MoTe_2$ remains intrinsic and exhibits photoluminescence (PL) emission and reflection peaks near 1.16 eV at 6.5 K (**Fig. S4b of the Supplemental Material)**, which we attribute to nearly degenerate neutral intralayer excitons $X_1$ and $X_2$. In the following discussion, we use X to denote these intralayer excitons collectively unless their layer pseudospin is specifically relevant. The close agreement between the PL and reflectance resonance energies indicates that bilayer $MoTe_2$ retains a direct optical bandgap, consistent with earlier reports[42-44]. We note that, although ab-initio theory predicts an electronic indirect gap with the conduction-band minimum at Q and valence-band maximum at K for bilayer $MoTe_2$[44], strong excitonic effects reshape the optical spectrum, yielding a direct optical bandgap.

Bilayer systems host multiple excitonic resonances with distinct layer characteristics, including intralayer, interlayer[19-21,35,36,45], and quadrupolar excitons[46,47]. In the field-dependent reflectance map (**Fig. 1b**), we observe a strong field-independent intralayer exciton, as well as two interlayer exciton species ($IX_1$ and $IX_2$) exhibiting a linear Stark shift. The PL of $IX_1$ and $IX_2$ is also observed (**Fig. S5)**. The voltage-derivative of the reflectance map reveals a clear avoided crossing between IX and the neutral intralayer exciton (**Fig. 1c**). Finally, we identify additional interlayer-like resonances, $X_Q$, which display weaker Stark shifts than IX.

We attribute $IX_1$ and $IX_2$ to momentum-direct interlayer excitons associated with the K–K transition. These interlayer excitons can hybridize with the intralayer neutral exciton X through interlayer carrier tunneling. Because electron tunneling is expected to be much weaker than hole tunneling, $IX_1$ ($IX_2$) primarily hybridizes with intralayer $X_1$ ($X_2$). A coupled-oscillator model including all four excitonic species captures the observed spectra well (**Fig. 1d**). From the fit, we extract an IX dipole moment of ~0.84 $e$·nm (see Methods), consistent with the interlayer spacing and indicative that the electron and hole are localized in opposite layers and with K-valley character. The extracted tunneling (hybridization) strength is on the order of 10 meV, in agreement with previous works[1,48,49].

Meanwhile, the $X_Q$ resonance shows a weaker Stark effect than the momentum-direct interlayer IX. One possible origin of $X_Q$ is a K–Q indirect transition, where stronger interlayer hybridization at Q reduces the out-of-plane dipole moment and suppresses the Stark response. Despite K-Q being the electronic band edge, this momentum-indirect K–Q exciton is expected to have a smaller binding energy than the intralayer K-K exciton, raising its total energy above that of X. Another possibility is that $X_Q$ corresponds to a quadrupolar exciton, formed from the hybridization of interlayer excitons[46].

## III. FERMI POLARONS AT ZERO DISPLACEMENT FIELD

Next, we study doping-dependent PL by tuning the charge carrier density while maintaining a zero displacement field. Under these conditions ($V_{BG} = \alpha V_{TG}$, where $\alpha$ is the thickness ratio of the top and bottom hBN layers; here $\alpha = 0.556$), the doped carriers are evenly distributed across the two layers. Upon doping, we observe emissions from attractive polarons, labeled as $AP^+$ and $AP^-$ (**Fig. 2a**). Interestingly, on the hole-doped side, an additional weak emission labeled $IP^+$ is also observed. On the electron-doped side, three $AP^-$ branches are resolved near zero doping, all of which redshift with increasing density. Notably, the highest-energy branch exhibits fine structure, with multiple closely spaced peaks near zero gate voltage (**Fig. S4c)**.

**Figures 2b** and **2c** show the doping-dependent reflectance contrast $(R - R_0)/R_0$ and its energy derivative $d(R/R_0)/dE$, respectively. On the hole-doped side, in addition to the $AP^+$ feature, the higher energy $IP^+$ feature becomes even more pronounced in the low doping region (**Fig. S4a**). On the electron-doped side, we observe a single $AP^-$ branch, aligning well with the highest-energy branch observed in PL. This correspondence suggests that the highest-energy PL branch arises from momentum-direct excitons dressed by electrons, whereas the two lower-energy PL branches likely involve momentum-indirect transitions enabled by phonon-assisted emission. Notably, this momentum-direct $AP^-$ feature exhibits a fine structure similar to the PL line shape (**Fig. S4c)**.

## IV. ELECTRIC-FIELD TUNING OF INTERLAYER POLARONS

We further investigate the rich spectroscopic features via doping-dependent PL and reflectance under asymmetric gating conditions, $V_{BG} = \alpha V_{TG} + \delta$, where the constant gate offset $\delta$ creates a finite out-of-plane electric field (**Fig. 3**). In the charge-neutral regime (region I), the excitonic features resemble the zero-field, zero-doping case **(Figs. 1 & 2)**. On the hole-doped side, however, two distinct regimes emerge, labeled as regions II and III. In particular, region II again features a prominent $IP^+$ resonance. This branch redshifts with increasing doping, while the $AP^+$ feature remains similar to that under symmetric gating. Upon entering region III, $IP^+$ turns into a blueshifting branch resembling the repulsive polaron $RP^+$, and an additional $AP^+$ feature appears. The gate-voltage extent of region II widens with increasing offset $\delta$ (**Figs. S6–S8**).

The excitonic behavior on the electron-doped side is more complex, but it also exhibits two distinct regimes. At low doping, we observe a repulsive polaron branch, $RP^-$, together with multiple $AP^-$ resonances. As the electron density increases, the spectra show a kink that marks a crossover into the high-doping regime. Beyond this crossover, the gate-dependent spectra again resemble those observed at zero displacement field. This richer behavior on the electron-doped side likely reflects the multivalley nature of the conduction band: electrons primarily occupy the Q valleys, but nearby K valleys of comparable energy may also become populated under a finite electric field. The coexistence of multiple Fermi seas, with different coupling to excitons in K valleys, can therefore generate a larger manifold of polaron states with fine structure, which we collectively denote as $AP^-$ (**Figs. 2** and **3**).

## V. THEORETICAL MODEL OF INTERLAYER POLARONS

To understand the doping- and electric-field-dependent behavior, we focus on the hole-doped regime, where holes occupy only the K valleys, making it less complex than the electron-doped regime. We attribute the lower-energy feature ($AP^+$) to the attractive intralayer polaron and the higher-energy feature ($IP^+$) to an interlayer polaron—a hybridized *intra*layer exciton dressed by the holes in the opposite layer.

At low hole density (region II), the electric field fully polarizes the holes into one layer, leaving the opposite layer undoped (inset of **Fig. 4b**). Excitons residing in the undoped layer can then interact with carriers in the adjacent layer through interlayer Coulomb interactions, forming a distinct polaron state. Notably, its exciton component is not entirely localized in the undoped layer but forms a coherent hybridization between intra- and interlayer excitons (**Fig. 4a**). Because the coupling between interlayer exciton and charge carriers is weaker than the intralayer exciton–carrier interaction, the $IP^+$ appears at a higher energy than $AP^+$. Furthermore, because the holes that form the polaron dressing of $IP^+$ reside in the opposite layer, the phase-space-filling blueshift[8,50]—that typically compensates for the attractive-interaction redshift in same-layer dressing—is largely absent, such that $IP^+$ exhibits a distinct redshift with increasing hole density.

As hole doping increases, the electric field can no longer maintain complete layer polarization, leading to a finite hole density in both layers (region III, inset of **Fig. 4b**). Excitons in either layer are then strongly dressed by the intralayer charge carriers, giving rise to a second attractive-repulsive polaron pair, $AP^+$ and $RP^+$, and a sudden blueshift of $IP^+$ that evolves continuously into the repulsive polaron.

We model the interlayer polaron with a field-theoretic framework based on the matrix Green's function of the bilayer exciton system (see the Supplemental Material). The key physical ingredients are the coherent interlayer hole tunneling, extracted from the intrinsic reflectance spectrum, the coupling of the excitons to the hole Fermi seas, and a hole-induced hybridization coupling of intra- and interlayer excitons (denoted as $\bar{g}_{X\text{-}IX}$ in **Fig. 4a**). We evaluate the self-energy within the non-self-consistent T-matrix approach[17,18,51], renormalizing the bare coupling via the vacuum T-matrix pole condition. This relates the coupling strength to two physical binding energies: the intralayer trion binding energy of ~20 meV, and the interlayer trion binding energy of ~10 meV. Adjusting the hole-induced hybridization $\bar{g}_{X\text{-}IX}$ and the theoretical Stark shift to match the experiment, the calculated absorption spectrum is in quantitative agreement with experiment (**Fig. 4b and c)**.

On the electron-doped side, multiple Fermi seas, together with layer polarization, can produce a larger manifold of polaron states, as discussed above. Even with this added complexity, our theoretical framework captures the key qualitative trends, where intralayer and interlayer polarons can still coexist when electrons are strongly layer-polarized. Upon entering the highly doped regime, we observe a kink and the two branches merge (see **Fig. S8**). A full account of the fine structure and its gate dependence is beyond the scope of this work and will be pursued in future studies.

## VI. CONCLUSION AND OUTLOOK

Our results establish bilayer $MoTe_2$ as a platform for quantum impurity physics in which both the impurity and the bath carry an internal degree of freedom, and in which the pseudospin and coherent coupling of the impurity states and the bath can be tuned independently. With electrical control over exciton energy, lifetime, and transport, one can envision on-demand trapping, guiding, and cooling of excitons[40,52,53], thereby opening an exciting path toward collective phases such as exciton condensates and superfluids[28-32], where the interaction-controlled layer coherence identified here may help stabilize interlayer coherence. A further extension is exciton–carrier dressing in correlated and topological phases of twisted $MoTe_2$. Because these phases are themselves governed by the layer pseudospin, intralayer and interlayer polaron formation may acquire distinct signatures depending on the underlying many-body order, offering spectroscopic access to fractional Chern insulators and their anyon excitations in twisted $MoTe_2$[23-27].

Carrier dressing in a valley-contrasting semiconductor introduces ingredients absent in conventional platforms: spin–valley locking and valley-selective optical selection rules, which can also couple to the layer pseudospin[15]. Harnessing the coherent interplay among the layer, charge, and valley degrees of freedom—especially in regimes where phonon-assisted processes are suppressed—could enable new approaches to encoding, controlling, and transducing quantum information in van der Waals materials.

Finally, the electrical control demonstrated here is appealing for active nanophotonic and polaritonic architectures, where switchable responses could enable reconfigurable light–matter coupling[54-58]. It is also relevant to nonlinear optics, where Fermi polarons already exhibit exceedingly large optical nonlinearities[49,59-61]. Here in a bilayer an electric field reshapes the polaron wavefunction and its out-of-plane dipole moment, providing a potential route to further enhance the gate-tunable nonlinearity for quantum-optical applications.

## ACKNOWLEDGMENTS

This research is supported by ARO W911NF2510066, NSF DMR-2145712, and OMA-2120757. The fabrication of samples is supported by the U.S. Department of Energy, Office of Science, Office of Basic Energy Sciences Early Career Research Program under Award No. DE-SC-0022885. This research used Quantum Material Press (QPress) of the Center for Functional Nanomaterials (CFN), which is a U.S. Department of Energy Office of Science User Facility, at Brookhaven National Laboratory under Contract No. DE-SC0012704. K.W. and T.T. acknowledge support from the CREST (JPMJCR24A5), JST and World Premier International Research Center Initiative (WPI), MEXT, Japan for hBN synthesis. E.D., M.W. and R.S. acknowledge support from the DFG (German Research Foundation) – ProjectID 273811115 – SFB 1225 ISOQUANT, and Germany's Excellence Strategy EXC 2181/1 - 390900948 (the Heidelberg STRUCTURES Excellence Cluster).

## AUTHOR CONTRIBUTIONS

Y.Z. and R.N. conceived the project. R.N. fabricated the samples and performed the experiments. S.T.L. and S.P. contributed to establishing the glovebox-based exfoliation and stacking platform and developing the associated fabrication workflow, and S.P. contributed to the development of the exfoliation process. J.J.S. and H.J. helped develop flake-mapping and motor-control software for glovebox-based sample preparation. L.G., R.M., and L.Z. assisted with general sample fabrication. Y.Z., L.G., R.M., and B.G. helped with optical measurements. S.J. designed and built the custom 2D materials glove box system and protocols with supervision from A.L.F. and A.T.H. T.T. and K.W. provided hexagonal boron nitride samples. Y.Z., E.D., M.W., and R.S. contributed to the theoretical interpretation of the data. Y.Z. and R.N. contributed to the data analysis. Y.Z., R.N., and E.D. wrote the manuscript with extensive input from all authors.

## Figures

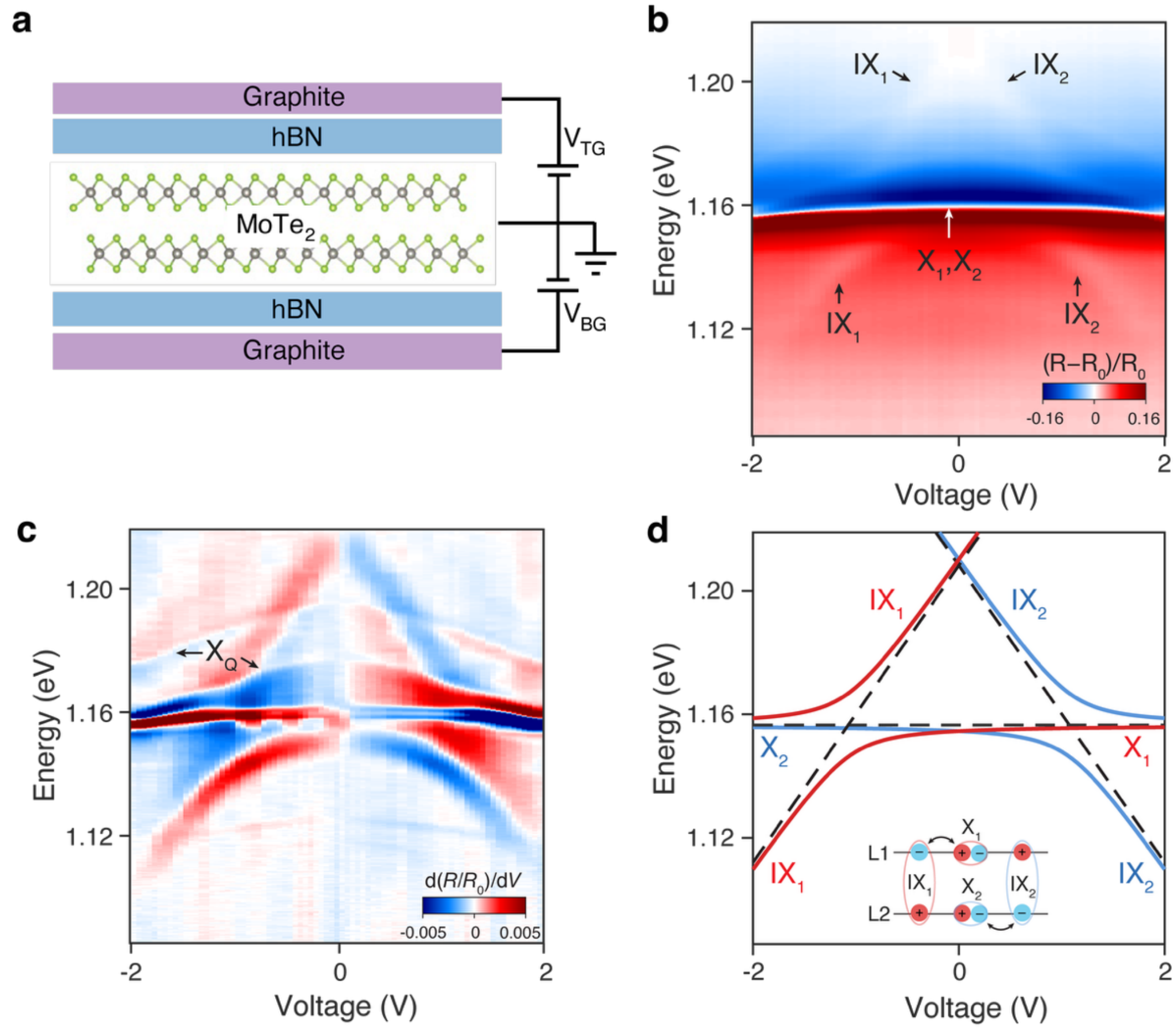

**FIG. 1. Intralayer and interlayer excitons in a $MoTe_2$ bilayer.** **a,** Schematic of a $MoTe_2$ bilayer heterostructure with dual, top and bottom-gate control. The bilayer $MoTe_2$ is encapsulated by two hBN multilayers, with thicknesses of 8 nm (top) and 4.5 nm (bottom). **b, c,** Electric-field-dependent reflectance contrast $(R - R_0)/R_0$, where $R_0$ is the baseline reflectance measured on the same spot at a highly hole-doped gate voltage (**b**) Reflectance contrast as a function of top gate voltage $V_{TG}$, and its voltage derivative (**c**) when the materials are nearly intrinsic: $V_{BG} = -0.556\ V_{TG}$ at $T = 6.5$ K. $X_1$ and $X_2$ denote intralayer K-valley excitons, $IX_1$ and $IX_2$ denote interlayer excitons with finite out-of-plane dipoles, and $X_Q$ denotes a Q-valley-related excitonic branch. **d,** Extracted electric-field-dependent exciton energy fitted using a coupled oscillator model. Inset: Illustration of intralayer and interlayer excitons in a $MoTe_2$ bilayer.

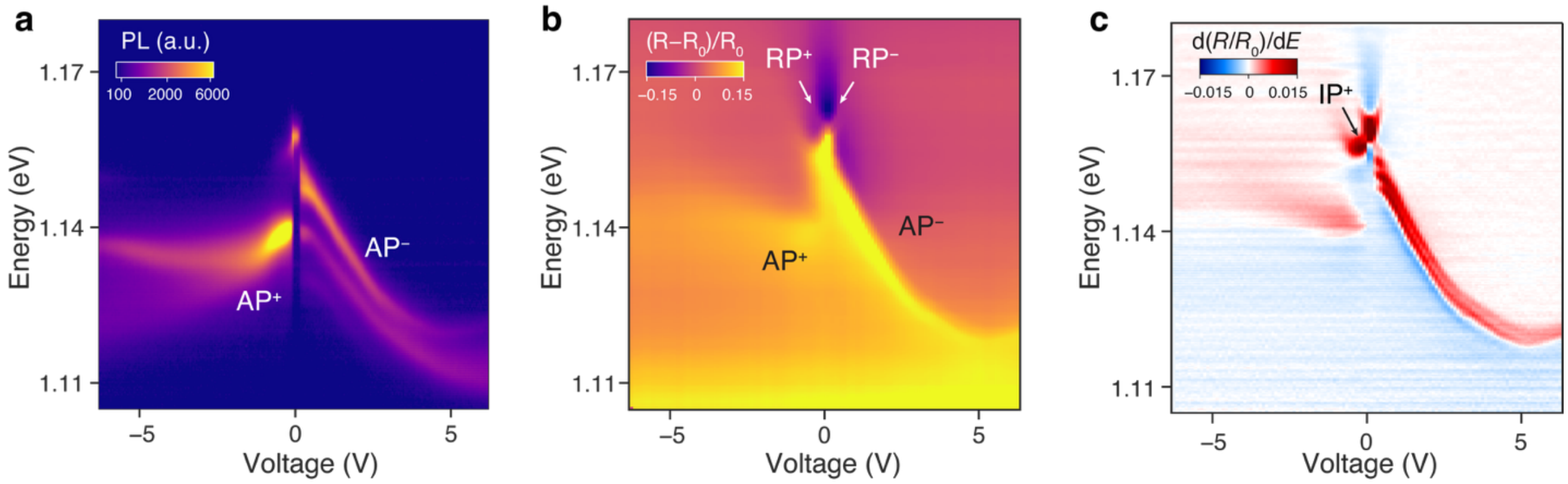

**FIG. 2. Doping-dependent exciton spectroscopy of a $MoTe_2$ bilayer under zero electric field. a**, Doping-dependent photoluminescence (PL) spectra under a 635 nm laser excitation as a function of top gate voltage ($V_{TG}$), with the gate voltage determined as $V_{BG} = 0.556 V_{TG}$. The PL intensity is shown using a square-root color scale to enhance the dynamic range. **b**, Doping-dependent reflectance contrast $(R - R_0)/R_0$ and **c**, its energy derivative. Here $R_0$ is the baseline reflectance measured from a nearby bare gold electrode (see Appendix B). $AP^{\pm}$ and $RP^{\pm}$ denote attractive and repulsive polarons, respectively, while IP+ denotes the interlayer polaron observed on the hole-doped side. The + (−) superscript indicates hole (electron) doping.

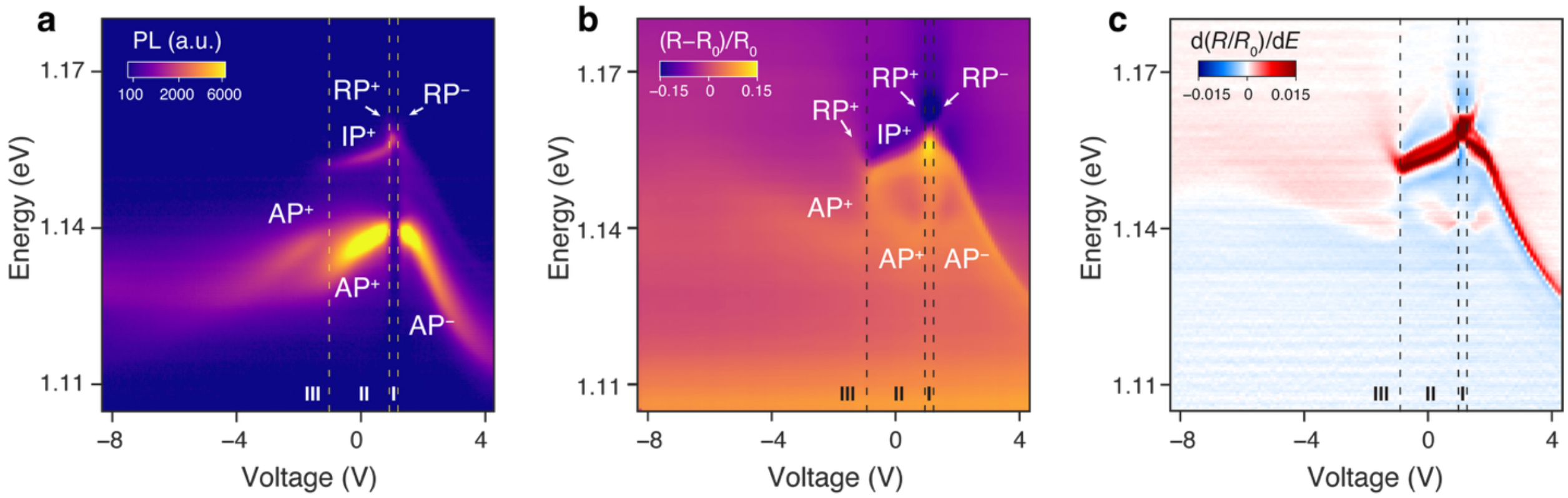

**FIG. 3. Doping-dependent exciton spectroscopy of a $MoTe_2$ bilayer under a finite electric field. a**, Doping-dependent PL spectra under a 635 nm laser excitation as a function of top gate voltage ($V_{TG}$), with the gate voltage determined as $V_{BG} = 0.556V_{TG} - 1.112$V. The PL intensity is shown using a square-root color scale to enhance the dynamic range. **b**, Doping-dependent reflectance contrast $(R - R_0)/R_0$ and **c**, its energy derivative. Here $R_0$ is the baseline reflectance measured on the gold electrode (see Appendix B). $AP^{\pm}$ and $RP^{\pm}$ denote attractive and repulsive polarons, respectively, while $IP^{+}$ denotes the interlayer polaron observed on the hole-doped side. The + (−) superscript indicates hole (electron) doping. Regions I–III correspond to the charge-neutral regime, the low-hole-density regime in which holes predominantly occupy one layer, and the higher-hole-density regime in which both layers are populated, respectively.

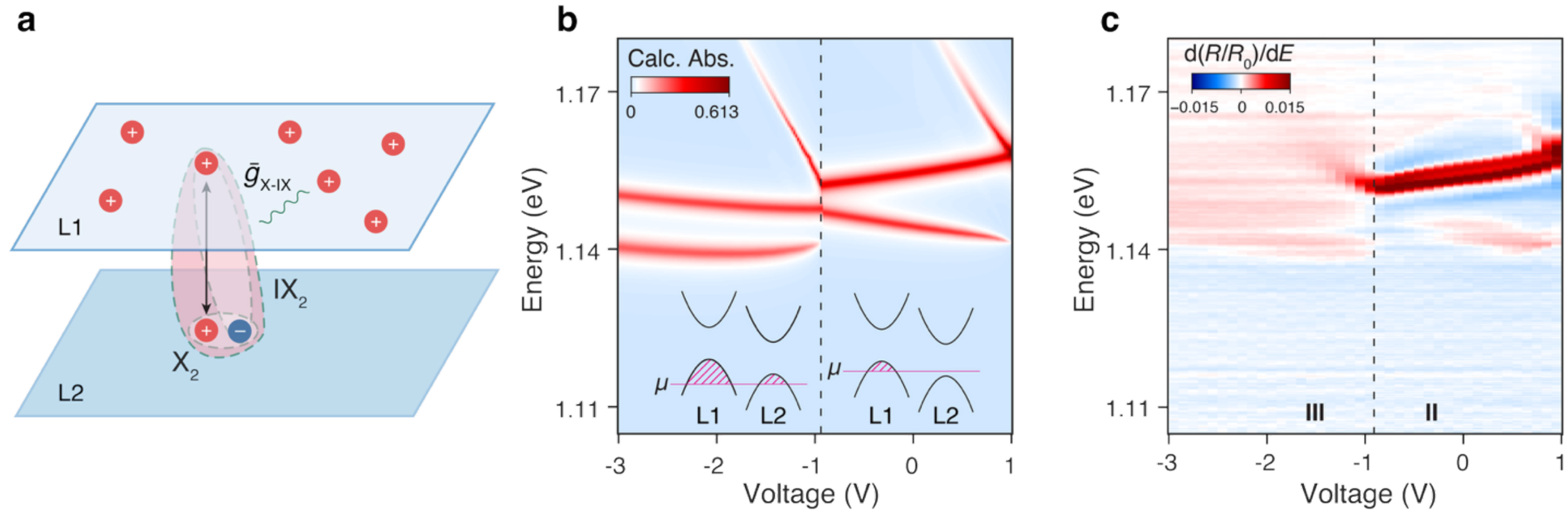

**FIG. 4. Theory of interlayer polarons in a $MoTe_2$ bilayer. a**, Illustration of a hybridized *intra*layer exciton between $X_2$ and $IX_2$ and its coupling to the hole Fermi sea in the opposite $MoTe_2$ layer. **b**, T-matrix-based theoretical calculation of the absorption spectrum. The insets show the band diagram and hole occupation of layer 1 (L1) and layer 2 (L2) under a finite electric-field while changing the chemical potential. In region II, only one layer is hole-doped, whereas in region III, both layers are doped. **c,** Experimental energy-derivative reflectance data under the same gating conditions show excellent agreement between theory and experiment.

## APPENDIX A: DEVICE FABRICATION

Graphite, 2H-$MoTe_2$ (HQ Graphene), and hBN flakes are mechanically exfoliated from bulk crystals onto a silicon substrate with a 285 nm $SiO_2$ layer. The silicon substrate is treated with oxygen plasma for 5 min before exfoliation. An optical microscope is used to identify hBN flakes for top- and bottom-gate dielectrics, as well as thin graphite flakes for gate electrodes, with their thickness estimated based on optical color contrast. Bulk 2H-$MoTe_2$ is exfoliated in an Ar-filled glove box, with water and oxygen levels less than 1 ppm. $MoTe_2$ flakes are identified based on color contrast under an optical microscope with a motorized stage. The heterostructure is subsequently assembled using a dry-transfer technique. Polydimethylsiloxane (PDMS) is squeezed and fixed by tape onto a glass slide, forming a dome shape. Polyvinyl chloride (PVC) tape is then attached to the dome. The PVC stamp picks up exfoliated flakes at ≈ 45 °C and releases them at ≈ 90 °C [62]. The full device is released onto a $SiO_2$/Si substrate. All fabrication steps involving $MoTe_2$, including exfoliation, optical identification, and stacking, are carried out in the same glovebox with an inert environment to prevent oxidation of $MoTe_2$. Then, electrical contacts are patterned by electron-beam lithography and lift-off patterning processes, with 5 nm of Cr and 90 nm of Au deposited by thermal evaporation.

## APPENDIX B: OPTICAL SPECTROSCOPY

The optical measurements are performed in the attoDRY800 cryostat with a custom-built confocal microscope. The base temperature of the cryostat is 6.5 K. The apochromatic objective lens equipped in the chamber has a numerical aperture of 0.82. The PL measurement is performed under excitation from both 635 nm and 980 nm diode lasers (Thorlabs). The reflectance measurement is performed using a halogen lamp (Thorlabs) as the light source. Reflectance contrast was calculated as $(R - R_0)\ /R_0$, where $R_0$ was obtained either at the same sample location under strong hole doping or from a nearby bare gold pad, as specified in the corresponding figure captions. The spectra are measured by a Horiba iHR320 monochromator using a 300 lines $mm^{-1}$ grating with a Synapse linear InGaAs arrays NIR detector.

## APPENDIX C: DUAL-GATE CALIBRATION AND VOLTAGE-SWEEP AXES

The gate dependence is presented as a function of the applied top and bottom gate voltages. This dual-gated geometry enables independent control of the carrier density and out-of-plane displacement field, in principle, modeled as a standard parallel-plate capacitor. However, because the analysis does not rely on absolute carrier density, we use experimentally applied gate voltages throughout the main text.

The relative gating efficiency of the top and bottom gates is calibrated experimentally from dual-gate PL maps. For each ($V_{TG}$ ,$V_{BG}$), the PL intensity was integrated over the spectral window of the intrinsic exciton. The integrated intrinsic-exciton PL intensity exhibits a diagonal feature in the ($V_{TG}$ ,$V_{BG}$) plane. This feature corresponds to gate-voltage combinations where the net electrostatic doping is minimized, reducing the density of gate-induced free carriers, suppressing charged-exciton or exciton-polaron formation, and thereby enhancing the neutral exciton emission. A linear fit to this feature gives the gate voltage ratio $V_{BG} = \alpha V_{TG}$, with $\alpha \approx 5/9$. This experimentally calibrated ratio provides a more accurate measure of the relative top- and bottom-gate efficiencies than the nominal hBN thickness ratio alone, as it directly accounts for device-specific electrostatics.

The calibrated gate voltage ratio was used to define sweep axes in the ($V_{TG}$ ,$V_{BG}$) plane. Gate trajectories parallel to $V_{BG} = \alpha V_{TG}$ predominantly tune the carrier density while maintaining a fixed out-of-plane displacement field. In particular, $V_{BG} = \alpha V_{TG}$ corresponds to an approximately zero-displacement-field doping scan, whereas offset trajectories $V_{BG} = \alpha V_{TG} + \delta$ were used to perform doping-dependent measurements at finite displacement fields. Conversely, gate trajectories parallel to $V_{BG} = -\alpha V_{TG}$ predominantly tune the out-of-plane displacement field while minimizing changes in electrostatic doping. The trajectory $V_{BG} = -\alpha V_{TG}$ corresponds to a displacement-field scan near charge neutrality, while offset trajectories $V_{BG} = -\alpha V_{TG} + \xi$ were used to perform displacement-field-dependent measurements at finite carrier density.

## APPENDIX D: ESTIMATION OF DOPING DENSITY AND ELECTRIC FIELD

The doping density and electric field were estimated from the applied gate voltage based on a parallel-plate capacitor model[52]. Thicknesses of the hBN (dielectric) layers are determined by atomic force microscopy (AFM), with thickness of top: $d_{TG}$ = 8 nm, $d_{BG}$ = 4.5 nm, respectively. The total doping density is calculated as $n = \frac{1}{e} \cdot (C_{TG} \cdot \Delta V_{TG} + C_{BG} \cdot \Delta V_{BG})$, where $C_{TG}$ and $C_{BG}$ are the capacitances of the top and bottom gates, respectively, and $\Delta V_{TG}$ and $\Delta V_{BG}$ are the applied gate voltages relative to the offset voltage to the band edge, respectively. The geometric capacitance is calculated using $C_{TG,BG} = \varepsilon_0 \varepsilon_{hBN} / d_{hBN}$. In the doping scan $V_{BG} = 0.556\ V_{TG} + \delta$, every +1 V increase of $V_{TG}$ (also increasing $V_{BG}$ by 0.556 V), increases the density by approximately $5.39 \times 10^{12}$ cm$^{-2}$. Assuming a twofold valley degeneracy and a parabolic valence band with a representative $MoTe_2$ hole effective mass, $m_h^* = 0.7 m_0$ [63], the hole chemical-potential shift is estimated as $E_F = \frac{2\pi\hbar^2 n}{g m_h^*}$, corresponds to a chemical-potential tuning rate of approximately 18 meV/V on the hole side.

The applied electric field can be calculated as $E = (V_{TG} - V_{BG})/(\frac{\varepsilon_{TMD}}{\varepsilon_{hBN}} \cdot d_{total} + d_{TMD})$ [52], where the finite thickness of the $MoTe_2$ bilayer is included because of the thin hBN dielectrics. Here for bilayer $MoTe_2$, $d_{TMD}$ ~ 1.4 nm, $d_{total}$ ~ 13.9 nm. For the dielectric constant we use $\varepsilon_{hBN}$ ~ 3.9, $\varepsilon_{TMD}$ ~ 7.2. In the electric field scan $V_{BG} = -\alpha V_{TG} + \xi$, every +1 V increase of $V_{TG}$ (decreases $V_{BG}$ by 0.556 V), increases the electric field by $\Delta E$= 0.0575 V/nm. From the fitting of Figure 1d, the slope of IX is extracted, ΔEnergy = 0.0481 eV/V. Using the estimated field conversion, the effective dipole moment is $p = \frac{\Delta Energy}{\Delta E}$ = 0.84 $e \cdot$ nm. This value is treated as an effective dipole moment because the field conversion relies on a simplified electrostatic model and representative dielectric constants.

## DATA AVAILABILITY

The data that support the findings of this study are available from the corresponding author upon reasonable request.